\documentclass[%
 reprint,
 amsmath,amssymb,
 aps,
]{revtex4-2}
\usepackage{braket}
\usepackage{color}
\usepackage{graphicx}
\usepackage{dcolumn}
\usepackage{bm}
\usepackage{ulem}
\usepackage{amssymb,amsmath}
\usepackage{physics}
\usepackage{subfigure}
\usepackage{mathtools}
\usepackage{hyperref} 
\usepackage{braket}
\usepackage{amsthm}

\begin{document}

\preprint{APS/123-QED}

\title{
Quantifying the Energy Relaxation Rate of Quantum States Using D-Wave Device and the Discovery of Long-Lived Multiqubit States
}

\author{Takashi Imoto}
\email{takashi.imoto@aist.go.jp}
\affiliation{Research Center for Emerging Computing Technologies,
National Institute of Advanced Industrial Science and Technology (AIST),
1-1-1 Umezono, Tsukuba, Ibaraki 305-8568, Japan.}

\author{Yuki Susa}  
\email{y-susa@nec.com}
\affiliation{Secure System Platform Research Laboratories, NEC Corporation, Kawasaki, Kanagawa 211-8666, Japan}
\affiliation{NEC-AIST Quantum Technology Cooperative Research Laboratory, National Institute of Advanced Industrial Science and Technology (AIST), Tsukuba, Ibaraki 305-8568, Japan}

\author{Ryoji Miyazaki}  
\email{miyazaki-aj@nec.com}
\affiliation{Secure System Platform Research Laboratories, NEC Corporation, Kawasaki, Kanagawa 211-8666, Japan}
\affiliation{NEC-AIST Quantum Technology Cooperative Research Laboratory, National Institute of Advanced Industrial Science and Technology (AIST), Tsukuba, Ibaraki 305-8568, Japan}

\author{Yuichiro Matsuzaki}
\email{ymatsuzaki872@g.chuo-u.ac.jp}
\affiliation{Department of Electrical, Electronic, and Communication Engineering, Faculty of Science and Engineering, Chuo University}

\date{\today}

\begin{abstract}
Quantum annealing has been demonstrated with superconducting qubits.  
Such a quantum annealer has been used to solve combinatorial optimization problems. Moreover, it serves as a quantum simulator for investigating the properties of quantum many-body systems.
However, the coherence properties of multi-qubit states provided by D-Wave Quantum Inc. have not been explored sufficiently. 
Here, using the D-Wave device,
we measure the energy relaxation rate of superconducting qubits and find long-lived multi-qubit states.
Specifically, we investigate the energy relaxation rate of the first excited states of a fully connected Ising model with a transverse field.
We find that the decay rate of the excited states of such a system with four qubits
is orders of magnitude smaller than that of the excited state of a single qubit, which demonstrates the existence of long-lived multi-qubit states.  
We elucidate the mechanism using an independent decoherence model that qualitatively describes the phenomenon. 
In addition, by using such a mechanism,
we theoretically predict a long-lived entangled state whose energy relaxation rate is smaller than that of the separable states. 
\end{abstract}

\maketitle


\section{Introduction}

Quantum annealing(QA), which has witnessed significant developments, is useful for solving combinatorial optimization problems
\cite{kadowaki1998quantum, farhi2000quantum, farhi2001quantum}.
D-Wave Quantum Inc. has developed a quantum annealing machine that consists of thousands of qubits \cite{johnson2011quantum}.
The combinatorial optimization problem can be mapped into a search for the ground state of the Ising Hamiltonian \cite{lucas2014ising, lechner2015quantum}.
Efficient clustering and machine learning using QA have been reported in the literature\cite{kurihara2014quantum, kumar2018quantum, amin2015searching, neven2008training, korenkevych2016benchmarking, benedetti2017quantum, willsch2020support, wilson2021quantum}.
Moreover, QA is useful for topological data analysis (TDA)\cite{berwald2018computing}.

Furthermore, a quantum annealing machine can be used not only for solving problems but also as a quantum simulator for quantum many-body systems in and out of equilibrium \cite{king2018observation, kairys2020simulating, harris2018phase, zhou2021experimental}.
There is considerable interest in the potential of D-Wave machines to exploit quantum properties, and recent experiments with a D-Wave machine have shown good agreement with theoretical predictions based on the Schr\"odinger dynamics \cite{king2022coherent, king2022quantum}. 
To this end, a study has attempted to use longitudinal magnetic fields to probe the dynamics of a qubit and distinguish the effects of noise in the D-Wave machine\cite{morrell2022signatures}.
Tunneling spectroscopy \cite{berkley2013tunneling} is a way to estimate the energy gap of the transverse-field Ising Hamiltonian, and numerical simulations of open quantum systems
have been performed to reproduce the results of the tunneling spectroscopy  
by using the D-Wave machine \cite{chen2020hoqst}.
When reverse quantum annealing was performed to search doubly degenerate ground states, the probability of finding one of the ground states was much larger than that of the other state, and this behavior was qualitatively reproduced by a theoretical model of decoherence \cite{bando2022breakdown}.

Quantum correlation plays an important role in a quantum simulator.
Experimental and theoretical results indicate that correlation among qubits lessens the impact of noise during QA \cite{weinberg2020scaling}.
Especially since an eigenstate of the Hamiltonian provided by the D-Wave machine can be entangled, it is important to study the coherence properties of entangled states.
The properties of the entanglement of the ground state during QA were experimentally and theoretically investigated
\cite{lanting2014entanglement,albash2015reexamination}.
However, the coherence properties of multi-qubit states generated by D-Wave have not been sufficiently explored.

In this paper, we measure the energy relaxation rate of the excited states during QA by using the D-Wave machine.
We employ a fully connected Ising model of four qubits for the problem Hamiltonian of QA. 
Further, we show that the energy relaxation rate of the first excited states of such a four-qubit system is much smaller than that of a single-qubit system.
We elucidate the mechanism of this phenomenon by using an independent decoherence model.
Moreover, by adopting such a mechanism, we theoretically predict the existence of long-lived entanglement whose energy relaxation rate is smaller than that of the separable states under the effect of independent decoherence.

\section{Experimental Results using D-Wave machine}
\subsection{$T_1$ Measurement Setup}
First, we explain the method for measuring the energy relaxation rate
of the excited states during QA by using a D-Wave machine.
Our method is similar to that used in Ref. \cite{harris2008probing} where an energy relaxation rate of a single rf SQUID is measured. 
On the other hand, our method allows us to measure the energy relaxation rate of the interacting qubits.
We perform the reverse quantum annealing (RQA) with a hot start where the initial state is the first excited state of the problem Hamiltonian, and we investigate 
the energy relaxation rate
of the state during RQA.
We remark that we use the D-Wave Advantage system 6.4 for our demonstration throughout this paper.
Although embedding techniques have been developed for a large number of qubits \cite{yang2016graph, okada2019improving, boothby2020next, cai2014practical, klymko2014adiabatic, boothby2016fast, zaribafiyan2017systematic}, we employ a fully connected Ising model with four qubits, which is the maximum system size without embedding.
Moreover, we use several varieties of qubits in the D-Wave machine.

\begin{figure}
    \centering
    \includegraphics[width=0.95\linewidth]{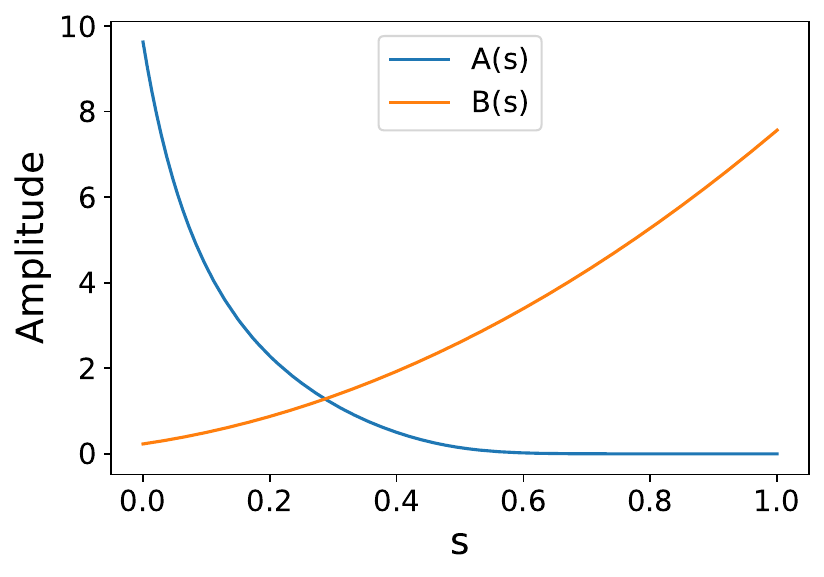}
    \caption{
    Scheduling functions $A(s)$ and $B(s)$ for the D-Wave device.
    This plot illustrates the amplitude of the scheduling functions $A(s)$ and $B(s)$ against the dimensionless parameter $s$, which ranges from $0$ to $1$.
    A(s) represents the amplitude of the transverse magnetic field, whereas $B(s)$ corresponds to the combined amplitude of the longitudinal magnetic field and Ising Hamiltonian.
    }\label{fig:amplitude_of_dwave}
\end{figure}

\begin{figure}
    \centering
    \includegraphics[width=1\linewidth]{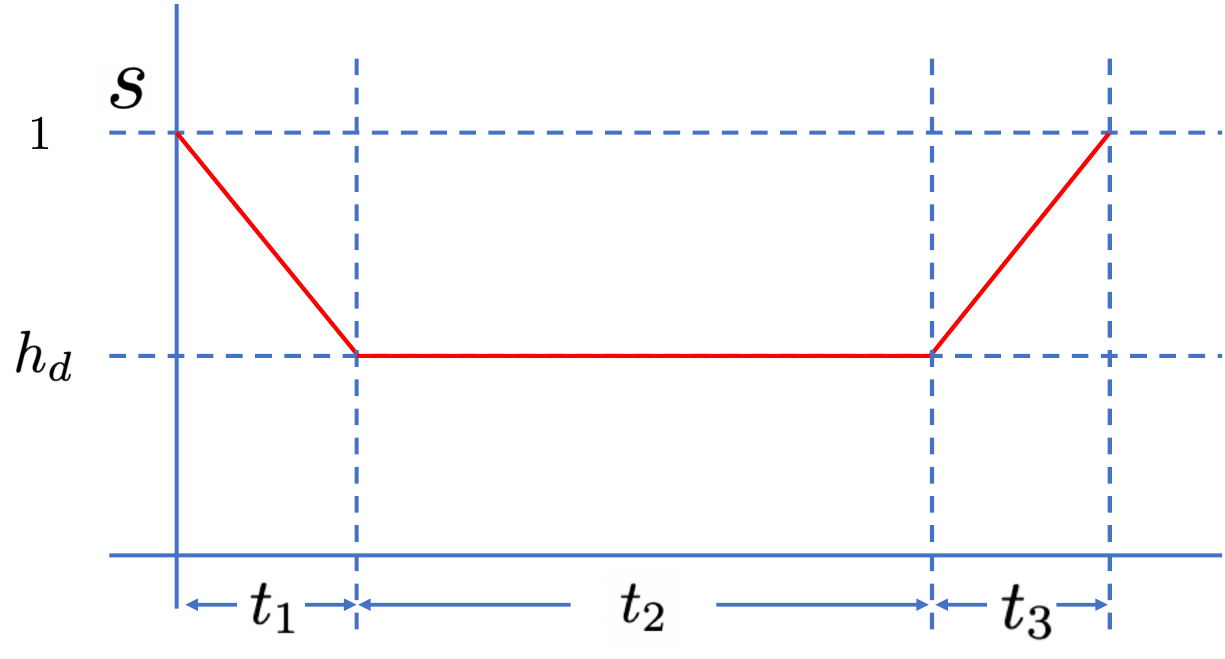}
    \caption{Schduling function of the reverse quantum annealing (RQA) against time $t$ to investigate energy relaxation rate in D-Wave Systems.
    The scheduling is divided into three distinct phases:$t_{1}$,$t_{2}$, and $t_{3}$.
    During $0<t<t_{1}$,$s$ linearly decreased from 1 to $h_{d}$.
    In $t_{1}<t<t_{2}$, $s$ is constant against time $t$.
    During $t_{2}<t<t_{3}$, $s$ linearly increases back to $1$, facilitating the measurement of the energy relaxation rate of the excited state.
    If we assume that non-adiabatic transitions during $0<t<t_{1}$ and $t_{2}<t<t_{3}$ are negligible, the relationship between the survival probability and $t_{2}$ can be used to determine the energy relaxation rate. 
    }
    \label{fig:t1_measurement_scheduling}
\end{figure}

The Hamiltonian is given as follows:
\begin{align}
    H(s)&=\frac{A(s)}{2}H_{D}+\frac{B(s)}{2}\biggl(g(s)H_{L}+H_{P}\biggr)\label{eq:gen_ann_ham}\\
    H_{P}&=-J\sum_{j<k}\sigma_{j}^{(z)}\sigma_{k}^{(z)}\\
    H_{L}&=\frac{h}{2}\sum_{j=1}^{N}\sigma_{j}^{(z)}\\
    H_{D}&=-B\sum_{j=1}^{N}\sigma_{j}^{(x)},
\end{align}
where $H_{P}$ denotes the problem Hamiltonian, $H_{D}$ denotes the drive Hamiltonian, $H_{L}$ denotes the Hamiltonian of the longitudinal magnetic field, $B$ ($h$) denotes the amplitude of the transverse (horizontal) magnetic fields, 
$J$ denotes the coupling strength, and $s$ denotes the time-dependent parameter for controlling the QA schedule. 
In this paper, we set $g(s)=1$.
The scheduling functions $A(s)$ and $B(s)$ for the D-Wave Advantage system 6.4 are given as shown in Fig. \ref{fig:amplitude_of_dwave}.
Throughout our paper, we set $h=1$ [GHz] and $B=1$ [GHz] unless specified otherwise.
Furthermore, when we consider a fully connected Ising model, we set $J=1$ and $N=4$.
By contrast, when we consider a non-interacting model (or a single-qubit model), we set $J=0$ and $N=1$ for simplicity.
We select the first excited state of the problem Hamiltonian as an initial state.
For the fully connected Ising model, the all-up state $\ket{\uparrow\uparrow\uparrow\uparrow}$ is the first excited state.
The schedule of the Hamiltonian is described as follows. (See Fig. \ref{fig:t1_measurement_scheduling})
First, from $t=0$ to $t=t_1$, we change the parameter of the Hamiltonian from $s=1$ to $s=h_{d}$ as a linear function of $t$, where we can control the strength of the transverse magnetic fields by tuning $h_d$.
Second, from $t=t_1$ to $t=t_1 + t_2$, we let the Hamiltonian remain in the same form at $s=h_{d}$.
Third, from $t=t_1 + t_2$ to $t=t_1 + t_2 +t_3$, we gradually change the parameter of the Hamiltonian from $s=h_{d}$ to $s=1$.
Fourth, we measure the state with a computational basis.
Finally, we repeat this protocol with various values of $t_2$.
We obtain the survival probability of the initial state after these processes.
We fit the survival probability by using $ce^{-\Gamma t_2}$ where $c$ is a constant factor, and we estimate the energy relaxation rate $\Gamma$.
Throughout this paper, 
we set $t_1=t_3=1$ $\mu$s, the annealing times $t_1$ and $t_3$ are set to be $1$ $\mu$s, and we define an energy relaxation time as the inverse of the energy relaxation rate.

\begin{figure*}[t!]
\includegraphics[width=180mm]{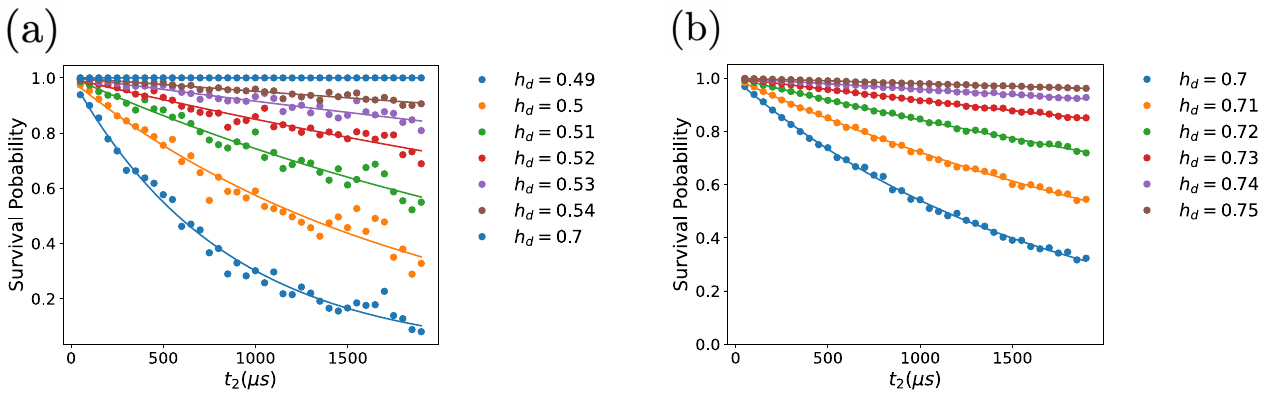}
\caption{(a)
    Survival probability plotted against $t_2$ for the fully connected Ising model ($J=-1$) with the transverse field. The solid line represents a fitted exponential decay curve.
    (b) Survival probability plotted against $t_2$ for the single qubit. 
    The solid line represents a fitted exponential decay curve.
    We set the fitting function to $a\exp(-t/T_{1})$ where $T_1$ is the energy relaxation time and $a$ is a constant factor.
    Each point is obtained with $100000$ measurements. Furthermore, we set $t_1=1\mu s$, $t_3=1\mu s$, and $h=1.0$.
}\label{fig:D-wave_result}
\end{figure*}

\subsection{Experimental Results}
\begin{figure}[h!]
    \centering
    \includegraphics[width=92mm]{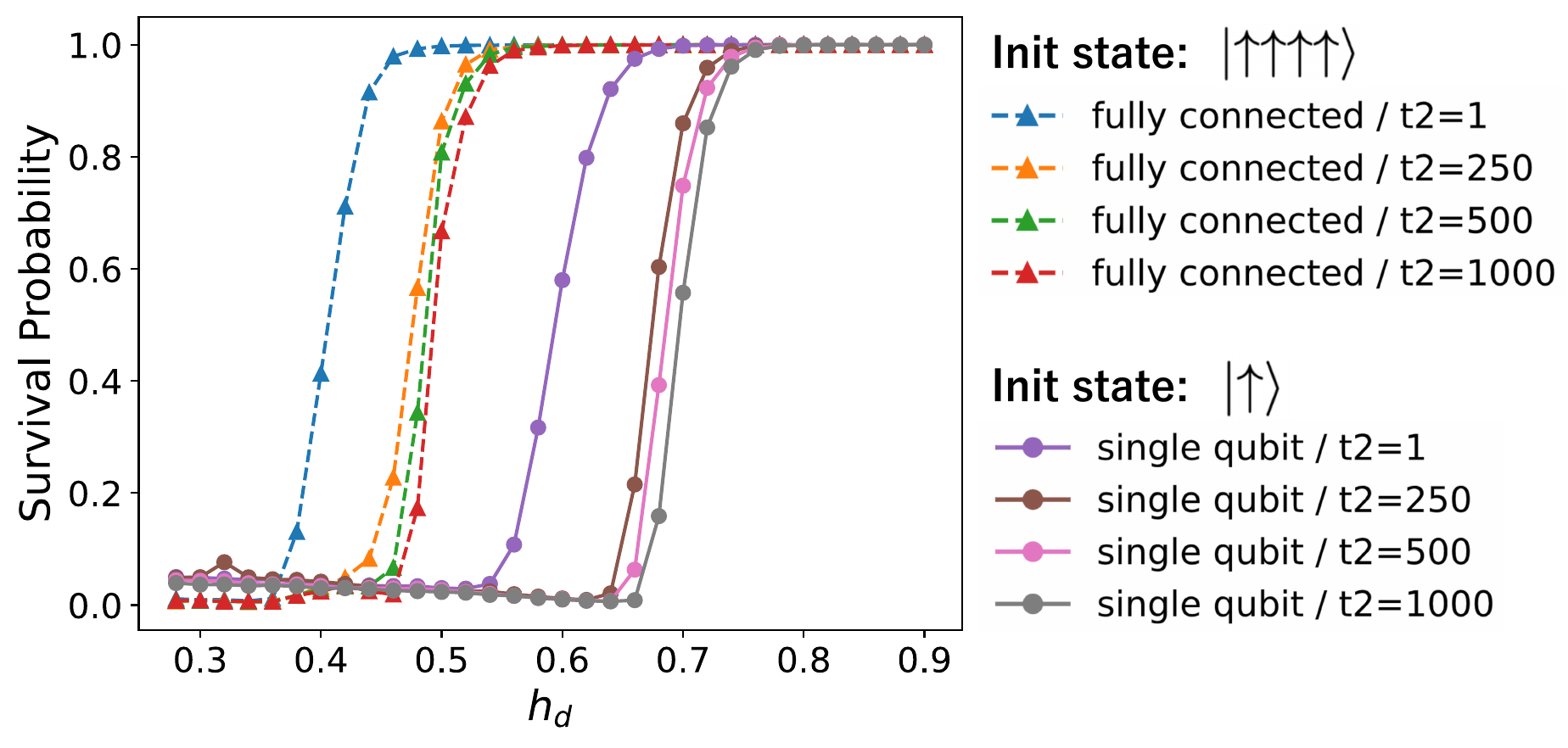}
    \caption{
    Survival probability plotted against $h_d$ for the fully connected Ising model ($J=-1$) with the transverse field and the single-qubit model.
    Each point is obtained with $100000$ measurements.
    Furthermore, we set $t_1=1\mu s$, $t_3=1\mu s$, and $h=1.0$.}\label{fig:s_prob_against_hd}
\end{figure}

Using the method described above, we measure the energy relaxation rate of the first excited state.
Moreover, we compare the energy relaxation rate of the state of the transverse-field Ising model with that of the single-qubit model.
Fig. \ref{fig:D-wave_result} (a) shows the survival probability plotted against $t_2$ in the case of the fully connected Ising model with the transverse field.
As can be seen, the survival probability decays exponentially.
Furthermore, as $h_d$ is decreased, the energy relaxation rate becomes larger.
Fig. \ref{fig:D-wave_result} (b) shows the survival probability plotted against $t_2$ in the case of the single-qubit model. 
Similar to the case of the fully connected Ising model, the energy relaxation rate becomes larger as $h_d$ is decreased. 
Importantly, the energy relaxation rate of the state of the single-qubit model is larger than that of the fully connected Ising model with the transverse field.
The details of the experimental data regarding the measured energy relaxation time are presented in Appendix \ref{sec:t1_table}.
To gain a deeper understanding of what occurs in the RQA, we perform numerical calculations in section \ref{sec:long-lived-numerical}.
We will discuss the origin of the long-lived multi-qubit state in section \ref{sec:perturbation_analysis}.
Additionally, the survival probability is plotted against $h_d$ in Fig. \ref{fig:s_prob_against_hd}.
Again, we confirm that the survival probability in the case of the fully connected Ising model is higher than that in the case of the non-interacting model.
These experimental results demonstrate that the energy relaxation time of the multi-qubit states can be longer than that of the single-qubit state.

Here, we discuss the possible origin of the decoherence in our experiments. 
For a superconducting flux qubit composed of three Josephson junctions \cite{chiorescu2003coherent,van2000quantum,mooij1999josephson}, the main source of decoherence is the change in the magnetic flux that penetrates the superconducting loop \cite{yoshihara2006decoherence,kakuyanagi2007dephasing}. This induces the fluctuation of the amplitude of $\hat{\sigma}_z$. 
Moreover, low-frequency flux noise was observed
for the RF-SQUID \cite{harris2008probing,lanting2009geometrical,harris2010experimental}, and the flux fluctuations induce the change in the amplitude of $\hat{\sigma}_z$.
From the aforementioned observations, we infer that the qubits in the D-Wave machine are also affected by the amplitude fluctuation of $\hat{\sigma}_z$.
Our results show that the multi-qubit states are robust against energy relaxation induced by such fluctuations.
To reveal the origin of the long-lived multi-qubit states, we consider the analysis with the Redfield master equation in the following subsection.

\subsection{Perturbative analysis for the long-lived qubit}\label{sec:perturbation_analysis}

\subsubsection{Analysis with the Redfield master equation}

To investigate the energy relaxation rate of the first excited state, we consider the Bloch-Redfield master equation defined as
\begin{align}
    \frac{d}{dt}&\rho(t)=-i[H(s(t)), \rho(t)]+\sum_{k,l}\sum_{\omega, \omega'}e^{i(\omega-\omega')}\gamma_{kl}(\omega')\notag\\
    &\times\biggl\{A_{l}(\omega')\rho(t)A_{k}^{\dag}(\omega)-A_{k}^{\dag}(\omega)A_{l}(\omega')\rho(t)\biggr\}+h.c
    \label{eq:redfield_eq}
\end{align}
where 
$\omega$ denotes the eigen energy difference of $H$, $\hat{A}_k$ denotes noise operator on the $k$-th qubit,
$A_{k}(\omega)=\sum_{\epsilon'-\epsilon=\omega}\ket{\psi_{\epsilon}}\bra{\psi_{\epsilon}}A_{k}\ket{\psi_{\epsilon'}}\bra{\psi_{\epsilon'}}$ denotes the noise operator associated with $\omega$, $\epsilon$ denotes the eigenvalue of the Hamiltonian $H(t)$, $\ket{\psi_{\epsilon}}$ denotes the eigenvector of the Hamiltonian $H(t)$, and $\gamma_{kl}(\omega)$ denotes the power spectrum density.
We set $\hat{\sigma}^{(z)}_{k}$ as the noise operator of $A_{k}$.
Also, we assume independent noise such as $\gamma_{kl}(\omega)\propto\delta _{k,l}$.

In the low-temperature limit, only transitions from the first excited state to the ground state need to be considered.
Consequently, noise operators can be expressed as
\begin{align}
    A_{k}(\omega)=&\ket{\phi_{gs}}\bra{\phi_{gs}}A_{k}\ket{\phi_{1st}}\bra{\phi_{1st}}\notag\\
    =&(\bra{\phi_{gs}}A_{k}\ket{\phi_{1st}})\ket{\phi_{gs}}\bra{\phi_{1st}}.
\end{align}

Here, $\ket{\phi_{gs}}\bra{\phi_{1st}}$ represents the transition operator from the first excited state to the ground state, while the transition matrix element $\bra{\phi_{gs}}A_{k}\ket{\phi_{1st}}$ characterized the strength of the transition.
Furthermore, incorporating the effect of the power spectral density from Eq. (\ref{eq:redfield_eq}), the relaxation rate can be evaluated as

\begin{align}
\tilde{\Gamma}=\sum_{j=1}^{L}\sqrt{\gamma_{j}(\omega)}|\bra{\phi_{gs}}\hat{\sigma}_{j}^{(z)}\ket{\phi_{1st}}|^2.    
\end{align}
We note that $\tilde{\Gamma}$ depends on the transition matrix and the energy gap between the ground state and the first excited state in the system Hamiltonian.

\subsubsection{Perturbative analysis for the long-lived qubit}

The decoherence rate depends on the power spectral density and the transition matrix such as $|\bra{\phi_{gs}}\hat{\sigma}_{j}^{(z)}\ket{\phi_{1st}}|$ as the previous section.
This suggests that the transition matrix plays a crucial role in determining the energy relaxation rate if the energy gap is the same.

To elucidate the mechanism of the low energy relaxation rate of multi-qubit states, we employ perturbation theory and obtain the analytical form of the ground state when the transverse magnetic field is small. 
Accordingly, the transition matrix provides a way to quantitatively evaluate the robustness of the first excited state against decoherence \cite{hornberger2009introduction}, based on the Redfield master equation.

We rewrite the total Hamiltonian (\ref{eq:gen_ann_ham}) as follows:
\begin{align}
    H=H_{0}+\lambda H_{1}.
\end{align}
where $H_{1}$ is the perturbative Hamiltonian and $H_{0}$ is the unpertabative Hamiltonian.
These Hamiltonians are defined by
\begin{align}
    H_{0}&=H_{P}+H_{L},\\
    H_{1}&=H_{D},
\end{align}
where we set the amplitude of the transverse magnetic field $B=1$.

Let us define $\hat{S}^{(a)}=\sum_{j=1}^{4}\hat{\sigma}_{j}^{(a)}\ (a=x,y,z)$ and $S^{(z)}$ as the eigenvalue of $\hat{S}_z$.
We consider the fully symmetric
representation corresponding to the maximum total spin, and the subspace is spanned by the Dicke states. In this subspace, we can specify the state by $S^{(z)}$.

\begin{figure}
\includegraphics[width=90mm]{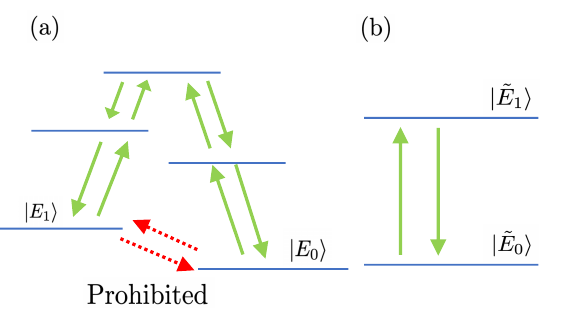}
\caption{(a) 
Energy level diagram of the fully connected Ising model with the transverse field.
For the small transverse magnetic field, we can use the lowest-order perturbation theory and obtain
$\ket{E_{0}}\approx\ket{\downarrow\downarrow\downarrow\downarrow}+\epsilon\sum_{j=1}^{4}\sigma_{j}^{(+)}\ket{\downarrow\downarrow\downarrow\downarrow}$ and $\ket{E_{1}}\approx\ket{\uparrow\uparrow\uparrow\uparrow}+\epsilon '\sum_{j=1}^{4}\sigma_{j}^{(-)}\ket{\uparrow\uparrow\uparrow\uparrow}$ where $\epsilon, \epsilon'$ denotes a small constant. 
In this case, the transition matrix elements of the local operators between these states are zero. 
This means that thus, decay from $\ket{E_{1}}$ to $\ket{E_{0}}$ is unlikely for the lowest order perturbation, which indicates a small energy relaxation rate
under the effect of local noise.
(b) 
Energy level diagram of a single qubit.
For some local operators, there are non-zero transition matrix elements between the ground state and the first excited state. 
In this case, decay can occur under the effect of local noise unless the noise direction is completely parallel to the quantization axis.
}
\label{fig:long-lived-fig}
\end{figure}

Using perturbation theory, we describe the $n$-th excited state as follows:
\begin{align}
    \ket{\phi_{n}}&=\ket{\phi_{n}^{(0)}}+\lambda\sum_{m\neq n}\frac{\bra{\phi_{m}^{(0)}}H_{D}\ket{\phi_{n}^{(0)}}}{E_{m}^{(0)}-E_{n}^{(0)}}\ket{\phi_{m}^{(0)}}
    +o(\lambda^{2}),\label{eq:purturbatid_state_formula}
\end{align}
where $\ket{\phi_{m}^{(0)}}$($E_{m}^{(0)}$) is the $m$-th 
energy eigenstate (eigenenergy) of $H_0$, and we assume that $E_m \leq E_{m'}$ for $m\leq m'$.
Here, we remark that $\ket{\phi_{m}^{(0)}}$ is the computational basis leading to
\begin{align}
   \bra{\phi_{n}^{(0)}}H_{L}\ket{\phi_{m}^{(0)}}=\bra{\phi_{n}^{(0)}}H_{P}\ket{\phi_{m}^{(0)}}=0\ \mbox{if}\ m\neq n
\end{align}
From Fig. \ref{fig:long-lived-fig} (a), we obtain the following eigenstates: 
\begin{align}
\ket{\phi_{0}^{(0)}}&=\ket{-4}\\
\ket{\phi_{1}^{(0)}}&=\ket{+4}\\
\ket{\phi_{2}^{(0)}}&=\ket{-2}\\
\ket{\phi_{3}^{(0)}}&=\ket{+2}\\
\ket{\phi_{4}^{(0)}}&=\ket{0}.
\end{align}
The corresponding energy eigenvalues are given by

\begin{align}
    E_{0}^{(0)}&=-8J-2h\\
    E_{1}^{(0)}&=-8J+2h\\
    E_{2}^{(0)}&=-2J-h\\
    E_{3}^{(0)}&=-2J+h\\
    E_{4}^{(0)}&=0.
\end{align}

Substituting these values into Eq. (\ref{eq:purturbatid_state_formula}), we obtain an explicit form of the ground state and the first excited state as follows:

\begin{align}
    \ket{\phi_{0}}&=\ket{-4}+\frac{\lambda}{6J+h}\ket{-2}+o(\lambda^{2})\\
    \ket{\phi_{1}}&=\ket{+4}+\frac{\lambda}{6J-h}\ket{+2}+o(\lambda^{2}).
\end{align}
Notably, these are entangled states.
The transition matrix between the ground state and the first excited state is given by
\begin{align}
    \bra{\phi_{0}}\hat{\sigma}_{j}^{(z)}\ket{\phi_{1}}=o(\lambda^{2})\label{eq:1qubit_transition_matrix_near_transverse_field}.
\end{align}
We can calculate the transition matrix of the other local operators. 
We obtain $\bra{\phi_{0}}\hat{\sigma}_{j}^{(x)}\ket{\phi_{1}}=o(\lambda^{2})$ and $\bra{\phi_{0}}\hat{\sigma}_{j}^{(y)}\ket{\phi_{1}}=o(\lambda^{2})$. 
Thus, this excited state is robust against other local noise as well.

Similarly, we calculate the eigenvector of the single-qubit model using perturbation theory for small transverse magnetic fields. 
The single-qubit model is given by
\begin{align} H_{single}=\hat{\sigma}^{(z)}+\lambda\hat{\sigma}^{(x)}.
\end{align}
We describe the eigenstates as follows:
\begin{align}    \ket{\psi_{0}}=\ket{\downarrow}+\lambda\ket{\uparrow}+o(\lambda^{2})\\
    \ket{\psi_{1}}=\ket{\uparrow}+\lambda\ket{\downarrow}+o(\lambda^{2}),
\end{align}
where $\ket{\psi_{0}}$($\ket{\psi_{1}}$) denotes the ground state (first excited state) of the single qubit.
The transition matrix element between the ground state and the first excited state is given by

\begin{align}
    \bra{\psi_{0}}\hat{\sigma}^{(z)}\ket{\psi_{1}}=o(\lambda)\label{eq:1qubit_transition_matrix_near_ising}.
\end{align}

By comparing Eqs. (\ref{eq:1qubit_transition_matrix_near_transverse_field}) and (\ref{eq:1qubit_transition_matrix_near_ising}), we confirm that the first excited state of the fully connected Ising model with the longitudinal and transverse fields is more robust against noise represented by $\hat{\sigma}_z$ compared to the single-qubit model.

Fig. \ref{fig:long-lived-fig} (a) and (b) illustrate the energy level diagram and the transition probability of the fully connected Ising model, highlighting the mechanism behind the robustness of the first excited state in our model.

Moreover, in the appendix \ref{sec:justification_redfield}, we qualitatively compare the experimentally observed energy relaxation rate with that predicted by the theoretical analysis calculated from the transition matrix. We conclude that such a theoretical analysis successfully reproduces the experimental results.

Our results differ from the decoherence-free subspace (DFS)\cite{lidar1998decoherence}. 
When the environment has a spatial correlation, it is possible to use logical qubits where the effect of the noise is suppressed. However, we cannot use the DFS for independent noise without spatial correlation. 
Our results described in this paper are useful for the independent noise model, and this is a crucial difference from the DFS.

Robust energy eigenstates to suppress an energy relaxation
were proposed\cite{dooley2018robust}.
These states are separable, and we can use these two energy eigenstates as a logical qubit, which is useful for quantum sensing \cite{dooley2018robust}.
Importantly, our results differ from the previous research because the energy eigenstates, which are robust against energy relaxation, are entangled in our case.

However, since we cannot perform the tomography with the D-Wave device, we cannot experimentally show the existence of the entanglement in our case. Instead, we will discuss possible long-lived entanglement as a theoretical proposal in the next section.

\section{Long-lived entangled state}\label{sec:long-lived-analysis}
In this section, we theoretically predict the existence of long-lived entangled states whose energy relaxation rate is smaller than that of a single qubit.
\subsection{Setup for Illustrating the Presence of an Entangled Long-Lived Qubit}
To theoretically demonstrate the presence of a long-lived entangled state in a noisy environment, we provide a detailed description of the model and the master equation that characterizes environmental effects, emphasizing the physical implications of the longitudinal and transverse fields in maintaining entanglement under decoherence.
To construct the entangled long-lived first excited state, we consider the fully connected Ising model with the longitudinal and transverse magnetic field $H(s)$ defined as
\begin{align}
    H(s)=(1-s)H_{D}+s\Bigl(H_{L}+H_{P}\Bigr)\label{eq:total_ham}
\end{align}
Here, the transverse magnetic field $H_{D}$, the longitudinal magnetic field $H_{L}$, and the fully connected Ising Hamiltonian $H_{P}$ are
defined as
\begin{align}
    H_{P}&=-J\sum_{j<k}\sigma_{j}^{(z)}\sigma_{k}^{(z)}\\
    H_{L}&=\frac{h}{2}\sum_{j=1}^{N}\sigma_{j}^{(z)}\\
    H_{D}&=-B\sum_{j=1}^{N}\sigma_{j}^{(x)}.
\end{align}
and $h$ denotes the amplitude of the longitudinal magnetic field.
We remark that the ground and first excited states of this Hamiltonian at $s=1$ are separable.
Let us define the case of a single qubit.
In this case, we assume $N=1$ for $N_{L}$ and $H_{D}$ with $H_{P}=0$.

In addition, to perform the numerical simulation of the Redfield master equation Eq. (\ref{eq:redfield_eq}), the low-temperature limit is assumed in this section; i.e., the power spectrum density $\gamma_{kl}(\omega)$ is constant
such as 
\begin{align}
\gamma_{kl}(\omega)=
   \begin{cases}
       \omega\eta\delta _{k,l}\ &(\omega> 0)\\
       0\ \ &(\omega\leq 0)
   \end{cases}\label{eq:psd_low_temp}
\end{align}
The initial state is set to be the first excited state of $H(s)$.
\\
\\
\subsection{Numerical calculation result}\label{sec:long-lived-numerical}

To investigate the energy relaxation rate, we simulate the Bloch-Redfield master equation Eq.(\ref{eq:redfield_eq}) from time $0$ to $T$ in the low-temperature limit.
Using the survival probability $|\braket{\phi_{init}}{\phi_{fin}(t)}|^{2}$ for each time $t$ and fitting the results to the function 
$f^{(N)}(t)=e^{-\alpha^{(N)}t}$
for $N=1,4$ where $\alpha^{(N)}$ is energy relaxation rate for $N$ qubit case, we obtain the energy relaxation rate of $s$ and $N$, $\ket{\phi_{init}}$ is the first excited state of $H(s)$ as the initial state and $\ket{\phi_{fin}(t)}$ is the final state via numerical time evolution using the Bloch-Redfield master equation Eq. (\ref{eq:redfield_eq}).

\begin{figure}
    \centering
    \includegraphics[width=1.0\linewidth]{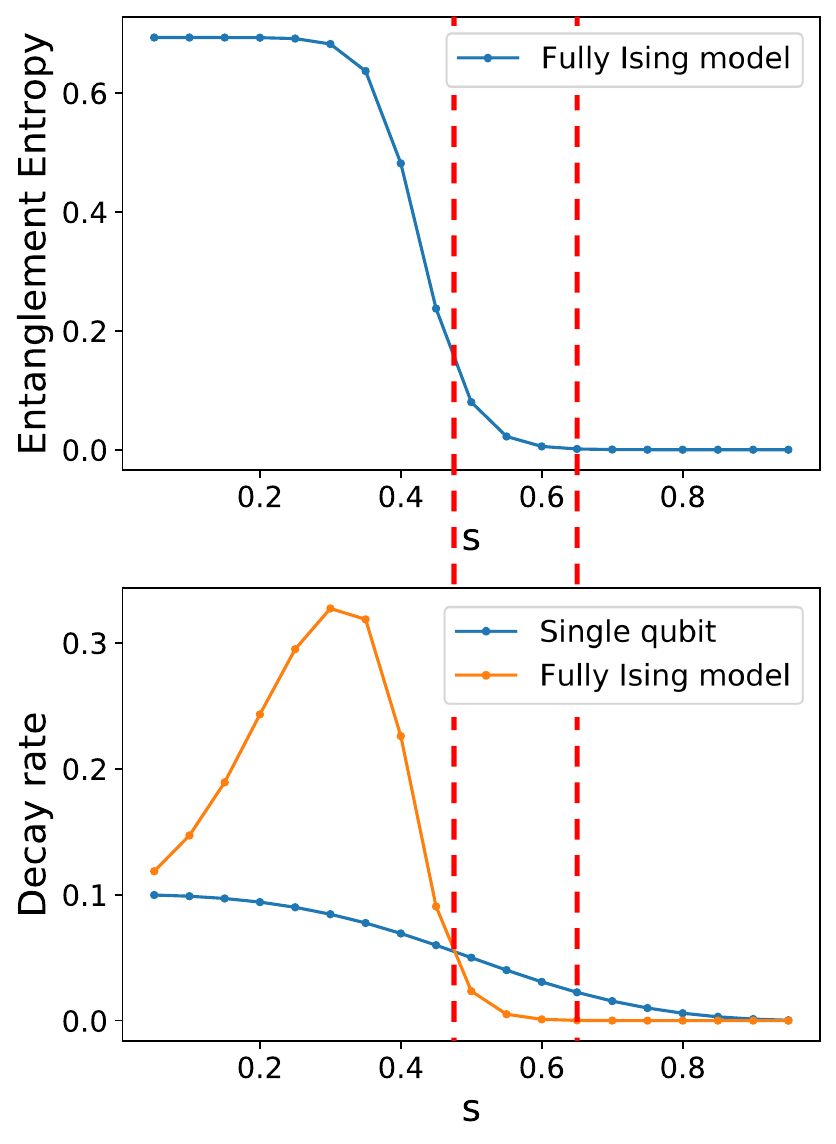}
    \caption{
    Entanglement entropy and energy relaxation rate are plotted against the parameter $s$ in the low-temperature limit and adopt the linear scheduling. 
    Namely, we use the following Hamiltonian $H(s)=(1-s)H_{D}+s(H_{L}+H_{P})$. We should note that, although s = 0, i.e. H(s) is the transverse field, the first excited state is a Dicke state that is entangled.
    The upper figure shows the entanglement entropy of the first excited state of the four-qubit fully connected Ising model against $s$. 
    On the other hand, the lower figure shows the energy relaxation rate of the fully connected Ising model and the single qubit against $s$ via solving the Bloch-Redfield master equation with an independent noise model.
    The lower figure shows the crossing point between these decay graphs at approximately $s\sim0.5$.
    Above this point, we can see that the energy relaxation rate of the single qubit case is larger than that of the fully Ising model case.
    In addition, this plot shows that the lifetime of the entangled state can be longer than that of the single qubit.
    }\label{fig:EE_decay}
\end{figure}

We derive the entanglement entropy of the fully connected Ising model Eq. (\ref{eq:total_ham}) for each $s$ in Fig. \ref{fig:EE_decay} of the upper figure by exact diagonalization.
The entanglement entropy is defined by $\Tr[\rho_{A}\log\rho_{A}]$, where $\rho_{A}$ denotes a reduced density matrix of the two qubits.
In addition, the energy relaxation rate of the fully connected Ising model and the single qubit are derived by simulating the master equation Eq. (\ref{eq:redfield_eq}) numerically.
In this simulation, we set the noise rate as $\eta=0.1$, time to simulate dynamics to derive the decay rate as $T=50$, the amplitude of the longitudinal magnetic field $h/2=0.01$, and the interaction constant $J=2$.
From the lower figure of Fig. \ref{fig:EE_decay}, the crossing point of the lines of the single qubit and the fully connected Ising model appears at $s\approx0.5$.
Below this point, the energy relaxation rate of the single qubit is larger than that of the fully connected Ising model.
Entangled states exist in this regime from the upper figure of Fig. \ref{fig:EE_decay}.
Therefore, we observe that the energy relaxation time of the entangled first excited state of the many-body model can be longer than the single qubit.

\section{Conclusion and future work}
In conclusion, we measured the energy relaxation rate of the first excited state during QA by using a D-Wave machine.
We found that the energy relaxation rate of the first excited state of the transverse-field Ising system with four qubits is much smaller than that of the single-qubit system.
The origin of the long-lived multi-qubit state is a small transition-matrix element of the noise operator between the ground state and the first excited state.
Moreover, we theoretically predict the existence of the long-lived entangled state whose decay rate is smaller than that of the single-qubit state.
Our results open new possibilities in the D-Wave machine to explore the coherence properties of the quantum many-body systems.

This work was supported by Leading Initiative for Excellent Young Researchers MEXT Japan and JST Presto
(Grant No. JPMJPR245B) Japan and CREST (Grant No. JPMJCR23I5). This paper is partly
based on results obtained from a project, JPNP16007,
commissioned by the New Energy and Industrial Technology Development Organization (NEDO), Japan.
This work was supported by JST Moonshot R\&D (Grant Number JPMJMS226C).

\appendix

\section{Energy relaxation time}\label{sec:t1_table}

As described in the main text, we measured the energy relaxation rate of the fully connected Ising model in the longitudinal and transverse fields as well as that of the single-qubit model.  
We define an energy relaxation time as the inverse of the energy relaxation rate.
The details of the energy relaxation time of the fully connected Ising model in the longitudinal and transverse fields (single-qubit model) are presented in TABLE. \ref{table:t1_and_hd_fully} (\ref{table:t1_and_hd_single}).
It is worth mentioning that the energy relaxation time of the single RF SQUID was measured in \cite{harris2008probing,harris2010experimental}. 
For the single RF SQUID, the energy relaxation time measured in our experiment is comparable with that measured in the previous results. However, the difference is that we measure the energy relaxation time of the excited state of multiple-qubits and compare it with that of the separable state
by using the D-Wave machine, which was not done before.

\begin{table}[hbtp]
  \caption{Energy relaxation time for each $h_d$ for the fully connected Ising model($J=1$)with the transverse field}\label{table:t1_and_hd_fully}
  \centering
  \begin{tabular}{cc}
    \hline
    $h_d$  & Energy relaxation time($\mu s$)  \\
    \hline \hline
    0.43  & 92.4  \\
    0.44  & 104.7   \\
    0.45  & 113.1  \\
    0.46  &  137.1  \\
    0.47 &  217.7   \\
    0.48  &  391.4  \\
    0.49  &  831.1  \\
    0.50  &  1831.6  \\
    0.51  &  3345.3  \\
    0.52  &  6232.7  \\
    0.53  &  11016.2  \\
    0.54  &  19732.0  \\
    0.55  &  33737.7  \\
    0.56  &  52823.3  \\
    0.57  &  110145.9  \\
    \hline
  \end{tabular}  
\end{table}

\begin{table}[hbtp]
  \caption{Energy relaxation time for each $h_d$ for the single qubit}\label{table:t1_and_hd_single}  
  \begin{tabular}{cc}
    \hline
    $h_d$  & Energy relaxation time($\mu s$)  \\
    \hline \hline
    0.70  & 1640.0  \\
    0.71  & 3088.4   \\
    0.72  & 5849.1   \\
    0.73  & 11571.8   \\
    0.74  & 23465.0   \\
    0.75  & 48150.8   \\
    \hline
  \end{tabular}\label{table:t1_and_hd_single}
\end{table}

\section{Validation of the Redfield master equation for D-Wave processor
}\label{sec:justification_redfield}
In this section, we analyze the energy relaxation rate $\Gamma$ using the Redfield master equation, which predicts:

\begin{align}
    \tilde{\Gamma}=\sqrt{\gamma(w_{eg})}\sum_{j=1}^{L}|\bra{\phi_{gs}}\hat{\sigma}_{j}^{(z)}\ket{\phi_{1st}}|^{2}\label{eq:redfield_analy}
\end{align}
where $\gamma(w)$ is the power spectrum density corresponding to the energy gap of $w$, $w_{eg}$ denotes the energy gap between the ground state and the first excited state, and $\ket{\phi_{gs}}$($\ket{\phi_{1st}}$) is the ground state (the first excited state).
From Eq.(\ref{eq:redfield_analy}) derived from the Redfield master equation, the energy relaxation rate should be the same if both the energy gap and transition matrix are the same. 
In this section, we measure the energy relaxation rate with the D-Wave devices for several parameters under the constraint that the energy gap and transition matrix are the same. 
Here, we calculate the energy gap and transition matrix by using the exact diagonalization with a classical computer based on the experimental parameters with the D-Wave device.
We will show that the measured energy relaxation rates are almost the same with such a constraint, and these results show that our theoretical analysis with the Redfield master equation is valid for reproducing the decoherence dynamics of qubits with the D-Wave devices.

We define the energy gap (defined as $\omega_{\mathrm{eg}}$)
and the transition matrix (defined as $S=\sum_{j=1}^{L}|\bra{\phi_{1st}}\hat{\sigma}_{j}^{(z)}\ket{\phi_{gs}}|^2$) from the Hamiltonian as
\begin{align}
    H(h_{d},h)=-&\frac{A(h_{d})}{2}\sum_{j=1}^{L}\hat{\sigma}_{j}^{(x)}\notag\\
    &+\frac{B(h_{d})}{2}\biggl(J\Bigl(\sum_{j=1}^{L}\hat{\sigma}_{j}^{(z)}\Bigr)^{2}+\frac{h}{2}\sum_{j=1}^{L}\hat{\sigma}_{j}^{(z)}\biggr)
\end{align}
where $A(h_{d})$ and $B(h_{d})$ denote scheduling function of D-Wave devices and $\frac{h}{2}$ denotes the amplitude of the longitudinal magnetic field.
We select multiple parameter sets such that the energy gap $\omega_{eg}$ and the transition matrix $S$ are nearly identical across systems with different numbers of qubits.
If the theoretical analysis with the Redfield master equation is valid, the measured energy relaxation rates with the parameter set should be approximately the same.  We adopt this approach to check the validity of using the Redfield master equation to explain the experimental results for each $k$.

First, we consider the two-qubit cases ($L=2$) for $h=0.5$ and $h_{d}=0.48,0.482,0.484,0.486,0.488,0.49$.
We derive the energy gaps and transition matrices of these Hamiltonians as shown in TABLE. \ref{tab:1q_2q_comparison}.

\begin{figure*}[t]
    \centering
    \includegraphics[width=1\linewidth]{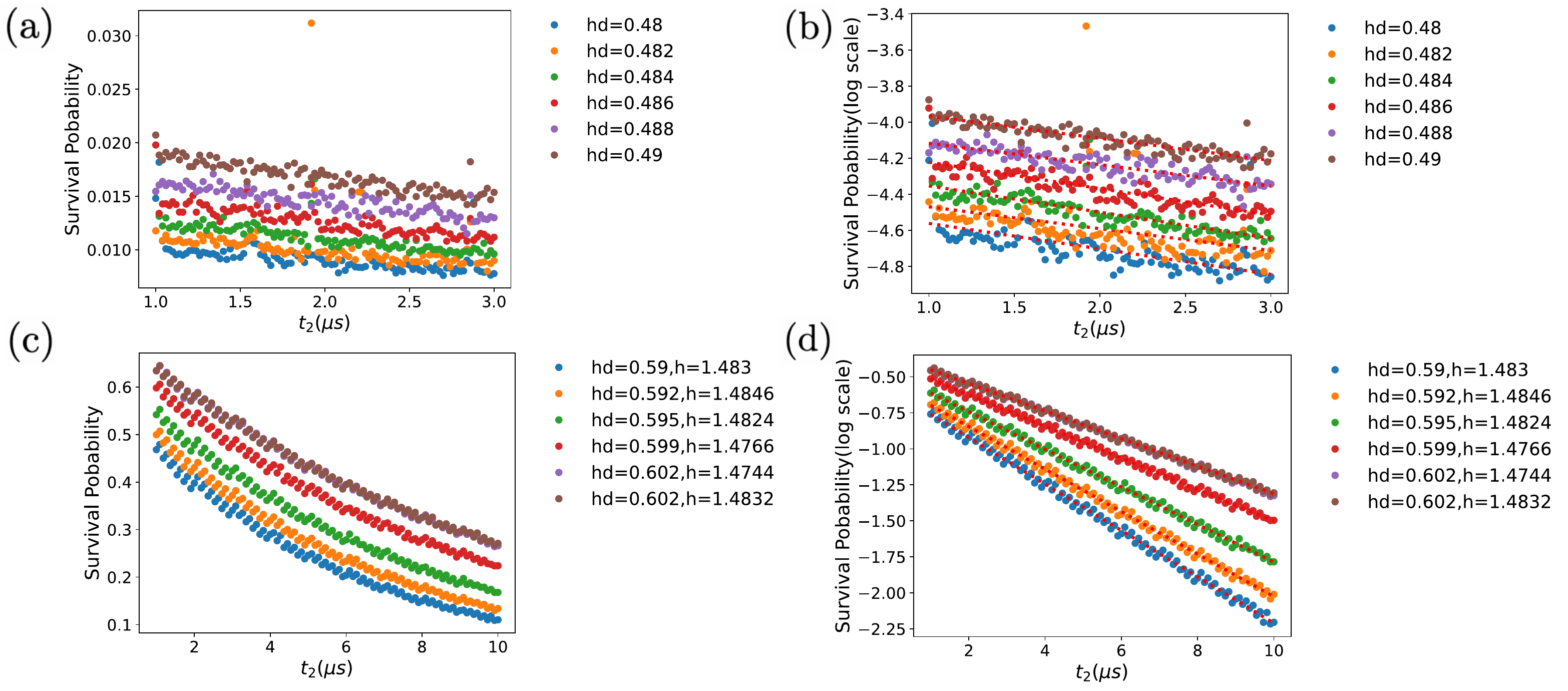}
    \caption{
    (a)Survival probability plotted against $t_{2}$ for the two-qubit model with the transverse field.
    (b)Survival probability plotted against $t_{2}$ for the two-qubit model with the transverse field in the log scale.
    From this plot, we derive the energy relaxation rate for several $h_{d}$ respectively as shown in TABLE.\ref{tab:1q_2q_comparison}.
    (c)Survival probability plotted against $t_{2}$ for the one-qubit model with the transverse field where the parameters of $h$ and $h_d$ are set as TABLE.\ref{tab:1q_2q_comparison}.
    We choose these parameters to have the same energy gap and transition matrix as those in the two-qubit case.
    (d)Survival probability plotted against $t_{2}$ for the one-qubit model with the transverse field in the log scale.
    From this plot, we derive the energy relaxation rate for several $h_{d}$ respectively as shown in TABLE.\ref{tab:1q_2q_comparison}.
    Throughout this figure, each point is obtained with 100000 measurements. Furthermore, we set $t_{1}=0.5\mu s$ and $t_{3}=0.5\mu s$.
    }
    \label{fig:one_two_qubit_survival_prob}
\end{figure*}
To experimentally obtain the energy relaxation rates of these cases, we plot the survival probability against $t_{2}$ for a two-qubit case in FIG. \ref{fig:one_two_qubit_survival_prob} (a) and (b).
From FIG. \ref{fig:one_two_qubit_survival_prob} (b) we derive the energy relaxation rate using the fitting function $\log P= at_{2}+b$ where $P$ is the probability.
These results are TABLE.\ref{tab:1q_2q_comparison}.

Second, we experimentally obtain the energy relaxation rate of the one-qubit models with the same energy gap and transition matrix as the two-qubit model.
The energy relaxation behavior of this experiment is shown in FIG. \ref{fig:one_two_qubit_survival_prob} (c) and (d).
From this plot, the decay rates for the one-qubit case corresponding to each of the two-qubit models are shown in TABLE.\ref{tab:1q_2q_comparison}.

These results show that our analysis to use the Redfield master equation is valid to explain the experimental results.

\begin{figure}
    \centering
    \includegraphics[width=1\linewidth]{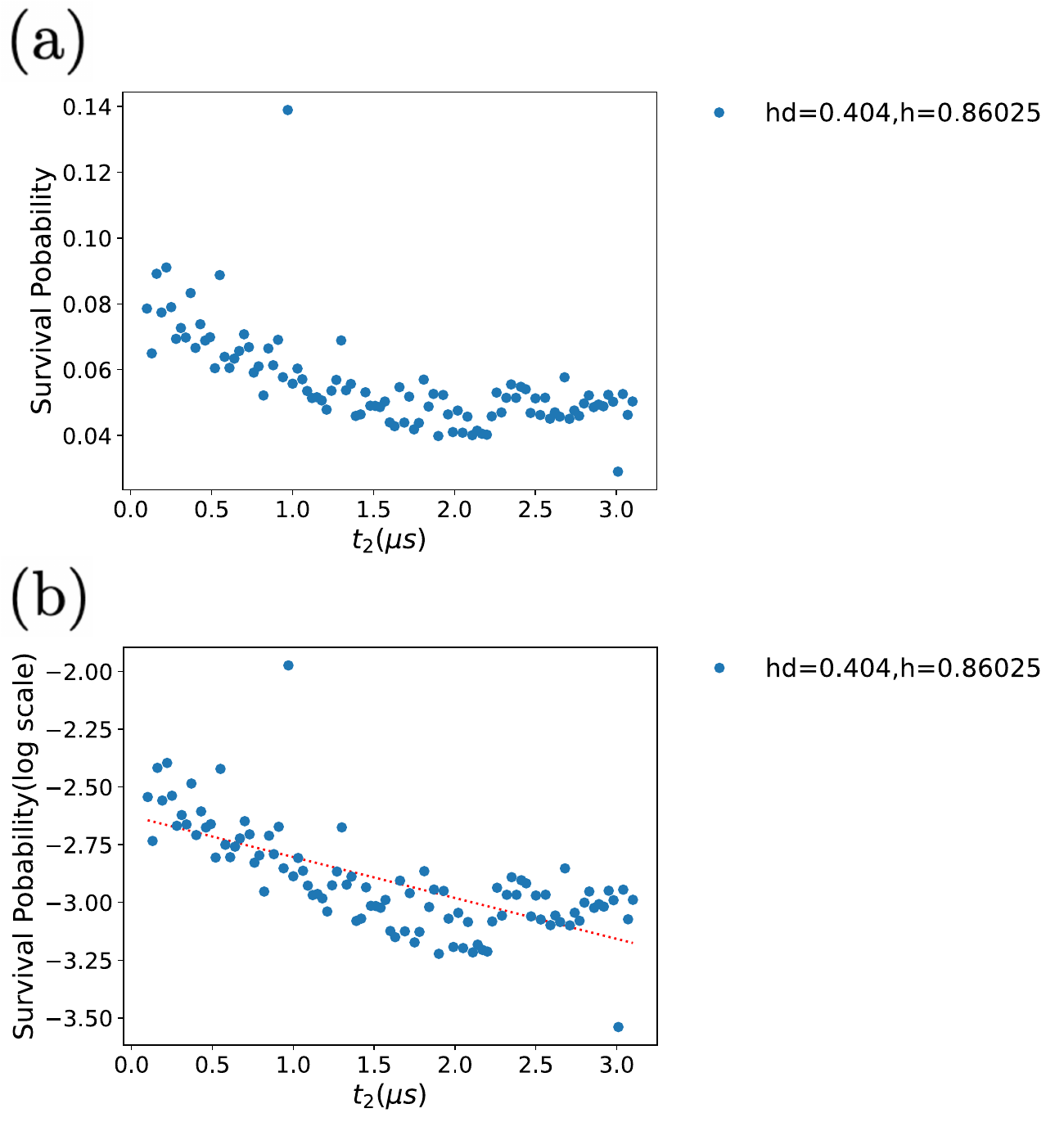}
    \caption{
    (a)Survival probability plotted against $t_{2}$ for the three-qubit model with the transverse field which is set to the two parameters $h$ and $h_{d}$ as 
    TABLE.\ref{tab:summary_comparison}
    to choose these parameters to have the same Energy gap and transition matrix as in the two-qubit case.
    (b)Survival probability plotted against $t_{2}$ for the three-qubit model with the transverse field on a log scale.
    From this plot, we derive the energy relaxation rate for several $h_{d}$ respectively as shown in TABLE.\ref{tab:summary_comparison}.
    Throughout this figure, each point is obtained with 1000000 measurements. Furthermore, we set $t_{1}=1\mu s$ and $t_{3}=1\mu s$.
    }
    \label{fig:three_qubit_survival_prob}
\end{figure}

\begin{figure}
    \centering
    \includegraphics[width=1\linewidth]{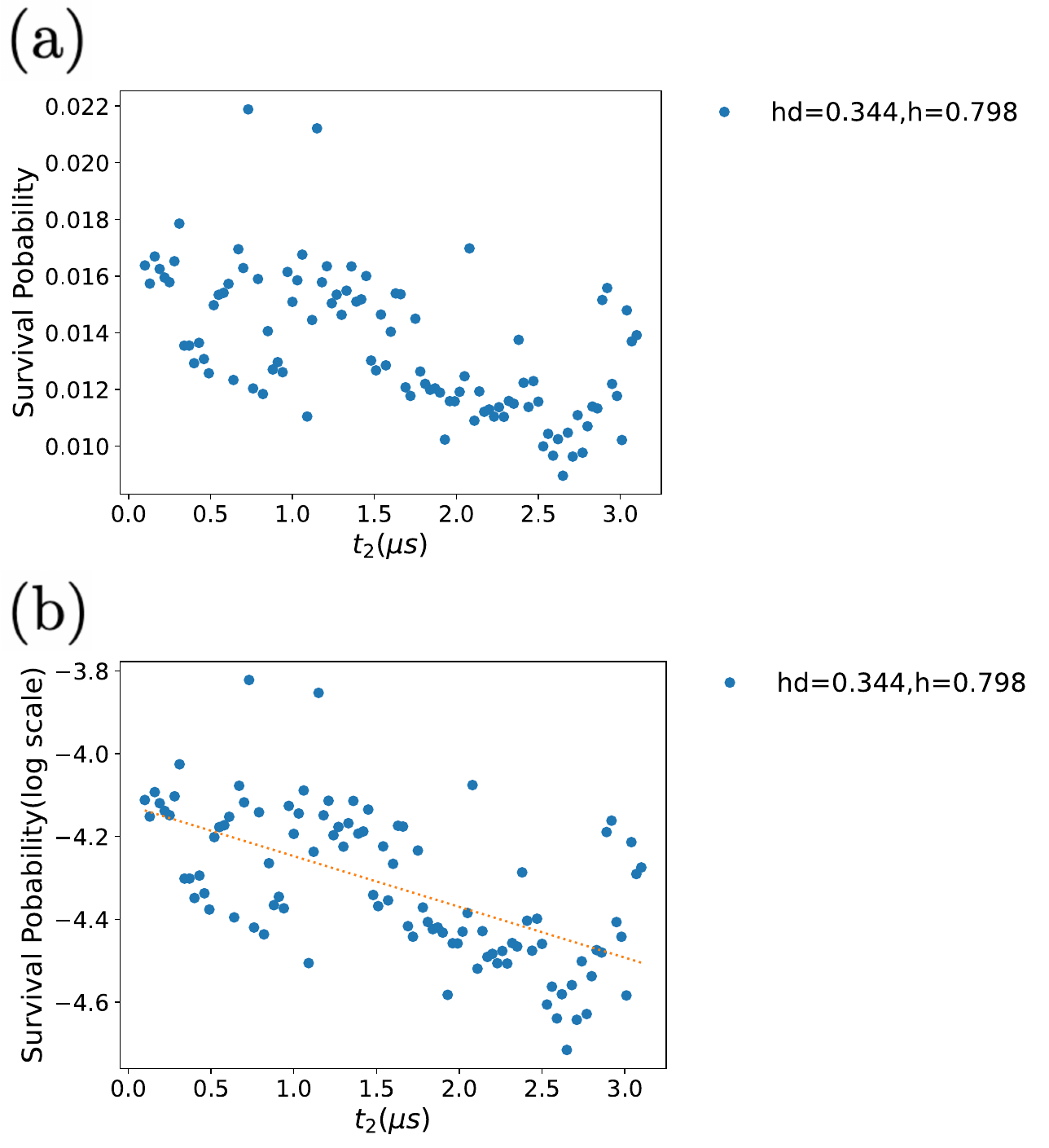}
    \caption{
    (a)Survival probability plotted against $t_{2}$ for the four-qubit model with the transverse field which is set to the two parameters $h$ and $h_{d}$ as 
    TABLE.\ref{tab:summary_comparison}
    to choose these parameters to have the same Energy gap and transition matrix as in the two-qubit case.
    (b)Survival probability plotted against $t_{2}$ for the four-qubit model with the transverse field in the log scale.
    From this plot, we derive the energy relaxation rate for several $h_{d}$ respectively as shown in     TABLE.\ref{tab:summary_comparison}.
    Throughout this figure, each point is obtained with 1000000 measurements. Furthermore, we set $t_{1}=1\mu s$ and $t_{3}=1\mu s$.
    }
    \label{fig:four_qubit_survival_prob}
\end{figure}

Furthermore, we experimentally obtain the energy relaxation rates of the three-qubit and four-qubit with the same energy gap and transition matrix as the two-qubit model($h_d=0.468$ and $h=0.5$).
We remark that to suppress the effect of the nonadiabatic transition and investigate the small value energy relaxation rate accurately, we set $t_{1}=t_{3}=1$ and $1000000$ measurements.
We plot the survival probability against $t_{2}$ for the three- and four-qubit case in FIG. \ref{fig:three_qubit_survival_prob} and FIG. \ref{fig:four_qubit_survival_prob}.
From the fitting, we obtain the energy relaxation rates in these cases in TABLE.\ref{tab:summary_comparison}.
It can be seen that these energy relaxation rates are close to the energy relaxation rate for $h_{d}=0.468$ in the case of the 2-qubit model.
Again, these results show that the Redfield master equation is valid to explain the experimental results.

\begin{table*}[ht]
\centering
\caption{
Comparison of energy relaxation rates between 1-qubit and 2-qubit systems with matched energy gap and transition matrix. 
Each row represents a pair of systems with nearly identical spectral properties about the energy gap and transition matrix.
Theoretical analysis based on the Redfield master equation predicts that the energy relaxation rate should be the same if the energy gap and transition matrix are the same. We experimentally measure the energy relaxation rate and validate such predictions.
The relaxation rates are derived from log-linear fits, hence the negative sign.}
\begin{tabular}{ccccc|ccccc}
\multicolumn{5}{c|}{\textbf{1-qubit system}} & \multicolumn{5}{c}{\textbf{2-qubit system}} \\
\hline
\hline
$h_d$ & $h$ & Energy Gap & Transition Matrix & Relaxation Rate $\Gamma$ &
$h_d$ & $h$ & Energy Gap & Transition Matrix & Relaxation Rate $\Gamma$ \\
\hline
\hline
0.590 & 1.483 & 2.46056  & $1.23547 \times 10^{-4}$ & -0.16110 &
0.480 & 0.5 & 2.46052 & $1.26994 \times 10^{-4}$ & -0.14117 \\
0.592 & 1.4846 & 2.47551 & $1.11799 \times 10^{-4}$ & -0.14764 &
0.482 & 0.5 & 2.47543 & $1.10151 \times 10^{-4}$ & -0.12121 \\
0.595 & 1.4824 & 2.49030 & $9.61010 \times 10^{-5}$ & -0.13046 &
0.484 & 0.5 & 2.49029 & $9.67579 \times 10^{-5}$ & -0.14539 \\
0.599 & 1.4766 & 2.50519 & $8.02154 \times 10^{-5}$ & -0.10955 &
0.486 & 0.5 & 2.50516 & $8.50286 \times 10^{-5}$ & -0.14639 \\
0.602 & 1.4744 & 2.51999 & $6.96026 \times 10^{-5}$ & -0.09663 &
0.488 & 0.5 & 2.52005 & $7.44975 \times 10^{-5}$ & -0.11854 \\
0.602 & 1.4832 & 2.53503 & $6.87792 \times 10^{-5}$ & -0.09575 &
0.490 & 0.5 & 2.53498 & $6.50641 \times 10^{-5}$ & -0.12294 \\
\hline
\end{tabular}
\label{tab:1q_2q_comparison}
\end{table*}

\begin{table*}[ht]
\centering
\caption{
Comparison of energy relaxation rates across systems with increasing qubit numbers.
All entries are chosen to have nearly identical energy gaps and transition matrix elements,
providing a consistent basis to test the predictive validity of the Redfield master equation.
}
\begin{tabular}{cccccc}
\hline
Qubits & $h_d$ & $h$ & Energy Gap & Transition Matrix & Relaxation Rate $\Gamma$ \\
\hline
1 (non-interacting) & 0.599 & 1.4766  & 2.50519  & $8.02154 \times 10^{-5}$ & -0.10955 \\
2 (coupled) & 0.486 & 0.5  & 2.50516 & $8.50286 \times 10^{-5}$ & -0.14639 \\
3 (coupled) & 0.404 & 0.86025 & 2.50510 & $8.69641 \times 10^{-5}$ & -0.17716 \\
4 (coupled) & 0.344 & 0.798   & 2.50464 & $8.31277 \times 10^{-5}$ & -0.12255 \\
\hline
\end{tabular}
\label{tab:summary_comparison}
\end{table*}

\nocite{*}

\bibliography{apssamp}

\end{document}